\documentstyle[12pt]{article}
\global\arraycolsep=2pt 

\newcommand{\leqsim}{\,\mbox{{\scriptsize $\stackrel{<}{\sim}$}}\,}
\newcommand{\heff}{{\cal H}_{{\rm eff}}(\Delta B=-1)}

\newcommand{\beq}{\begin{equation}}
\newcommand{\eeq}{\end{equation}}
\newcommand{\bea}{\begin{eqnarray}}
\newcommand{\eea}{\end{eqnarray}}
\newcommand{\VmA}{\mbox{{\scriptsize V--A}}}

\input{epsf}

\begin{document}

\thispagestyle{empty}

\rightline{TTP97-17}
\rightline{hep-ph/9704423}
\rightline{April 1997}
\bigskip
\boldmath
\begin{center}
{\Large{\bf Constraining the CKM angle $\gamma$ and}}

\vspace*{0.3truecm}

{\Large{\bf penguin contributions through combined}}

\vspace*{0.3truecm}

{\Large{\bf $B \to \pi K$ branching ratios}}
\end{center}
\unboldmath
\smallskip
\begin{center}
{\large{\sc Robert Fleischer}}\footnote{Internet: {\tt 
rf@ttpux1.physik.uni-karlsruhe.de}}\, and \, {\large{\sc Thomas 
Mannel}}\footnote{Internet: {\tt 
tm@ttpux7.physik.uni-karlsruhe.de}}\\
\vspace{0.3cm}
{\sl Institut f\"{u}r Theoretische Teilchenphysik}\\
{\sl Universit\"{a}t Karlsruhe}\\
{\sl D--76128 Karlsruhe, Germany}\\ 
\vspace{0.3cm}
\end{center}
\begin{abstract}
\noindent
Motivated by recent CLEO measurements of $B\to\pi K$ modes, we investigate 
their implications for the CKM angle $\gamma$ and a consistent description
of these decays within the Standard Model. Interestingly it turns out that 
already the measurement of the combined branching ratios $B^\pm\to\pi^\pm K$ 
and $B_d\to\pi^\mp K^\pm$ allows to derive stringent constraints on $\gamma$ 
which are complementary to the presently allowed range for that angle. This 
range, arising from the usual fits of the unitarity triangle, is typically 
symmetric around $\gamma = 90^\circ$, while our method can in principle 
{\it exclude} a range of this kind. Consistency within the Standard Model 
implies furthermore bounds on the ratio $r\equiv|T'|/|\tilde P|$ of the 
current-current and penguin operator contributions to $B_d\to\pi^\mp K^\pm$, 
and upper limits for the CP-violating asymmetry arising in that decay. 
Commonly accepted means to estimate $r$ yield values at the edge of 
compatibility with the present CLEO measurements.
\end{abstract}

\section{Introduction}
Recently the CLEO collaboration has presented a first measurement 
of some exclusive $B \to\pi K$ modes \cite{CLEO}. These modes are of 
particular interest since during the past years several strategies 
\cite{F97} have been proposed to use such decays for the extraction 
of angles of the unitarity triangle \cite{ut} of the CKM matrix \cite{ckm}, 
in particular for the angle $\gamma$ which is an experimental challenge
at $B$-factories. To this end flavor symmetries of
strong interactions are used. Unfortunately electroweak penguins play
in certain cases an important role and even spoil some of these methods
\cite{dh,GHLR-ewp}. Because of this feature rather complicated 
strategies~\cite{F97,GHLR-ewp,dh-gamma} are needed that are in most cases -- 
requiring e.g.\ the geometrical construction of quadrangles among 
$B\to\pi K$ decay amplitudes -- very difficult from an experimental 
point of view. 

A much simpler approach to determine $\gamma$ was proposed
in \cite{PAPIII}. It uses the branching ratios for the decays 
$B^+\to\pi^+K^0$, $B^0_d\to\pi^-K^+$ and their charge-conjugates. If the 
magnitude of the current-current amplitude $T'$ contributing to 
$B^0_d\to\pi^-K^+$ is known (we will discuss this point in more detail
later), two amplitude triangles can be constructed with the help of these 
branching ratios that allow in particular the extraction of $\gamma$.  

Since experimental data for $B^+\to\pi^+K^0$ and $B^0_d\to\pi^-K^+$ is
now starting to become available, we think it is an important and 
interesting issue to analyze the implications of these measurements for 
$\gamma$ and the 
description of these decays within the Standard Model in general. 
So far the CLEO collaboration has presented only results for the
combined branching ratios 
\begin{eqnarray}
\mbox{BR}(B^\pm\to\pi^\pm K)&\equiv&\frac{1}{2}\left[\mbox{BR}(B^+\to\pi^+K^0)
+\mbox{BR}(B^-\to\pi^-\overline{K^0})\right]\label{BR-char}\\
\mbox{BR}(B_d\to\pi^\mp K^\pm)&\equiv&\frac{1}{2}
\left[\mbox{BR}(B^0_d\to\pi^-K^+)
+\mbox{BR}(\overline{B^0_d}\to\pi^+K^-)\right]\label{BR-neut}
\end{eqnarray}
with rather large uncertainties:
\begin{eqnarray}
\mbox{BR}(B^\pm\to\pi^\pm K)&=&\left(2.3^{+1.1+0.2}_{-1.0-0.2}
\pm0.2\right)\cdot 
10^{-5}\label{BR-charres}\\
\mbox{BR}(B_d\to\pi^\mp K^\pm)&=&\left(1.5^{+0.5+0.1}_{-0.4-0.1}
\pm0.1\right)\cdot
10^{-5}\,.\label{BR-neutres}
\end{eqnarray}
At first sight one would think that measurements of such combined
branching ratios are not useful with respect of constraining $\gamma$. 
However, as we will work out in this paper, this is not the case. First,
non-trivial bounds on $\gamma$ of the structure
\begin{equation}\label{gamma-bound1}
0^\circ\leq\gamma\leq\gamma_0\quad\lor\quad180^\circ-\gamma_0\leq\gamma
\leq180^\circ\,,
\end{equation}
where $\gamma_0$ is related to the ratio of the combined branching ratios
(\ref{BR-char}) and (\ref{BR-neut}), can be obtained. Second, the ratio 
$r$ of the current-current amplitude $|T'|$ to the penguin amplitude 
$|\tilde P|$ contributing to $B_d\to\pi^\mp K^\pm$ can be constrained. 
Moreover it is possible to derive a simple formula for the maximally 
allowed value of the magnitude of the direct CP-violating asymmetry  
\begin{equation}\label{def-CP}
{\cal A}_{\rm CP}^{\rm dir}(B_d^0\to\pi^-K^+)\equiv\frac{\mbox{BR}
(B^0_d\to\pi^-K^+)-
\mbox{BR}(\overline{B^0_d}\to\pi^+K^-)}{\mbox{BR}(B^0_d\to\pi^-K^+)+
\mbox{BR}(\overline{B^0_d}\to\pi^+K^-)}
\end{equation}
that can be accommodated within the Standard Model. In the future, when the
CLEO measurements will become more accurate, these constraints on $\gamma$,
$r$ and $|{\cal A}_{\rm CP}^{\rm dir}(B_d^0\to\pi^-K^+)|$ should become 
more and more restrictive. If the amplitude ratio $r$ should lie far off its 
Standard Model expectation and CLEO should measure a CP-violating asymmetry 
in $B_d\to\pi^\mp K^\pm$ that is considerably larger than the corresponding
bounds on that observable obtained along the lines proposed in our paper, 
one would have indications for physics beyond the Standard Model. 

In Section~\ref{formulae} we set the stage for our discussion by giving
the formulae for the decay amplitudes of $B^+ \to\pi^+ K^0$ and 
$B^0_d \to\pi^- K^+$ within the Standard Model. Quantitative estimates for 
the branching ratios of these decays are presented in Section~\ref{estimates}.
There we also emphasize the importance of penguins with internal 
charm-quarks to get results of the same order of magnitude as the recent
CLEO measurements. The formula for $\gamma_0$ constraining $\gamma$ through 
(\ref{gamma-bound1}) is then derived in Section~\ref{bounds}, where we 
also give analytical expressions for bounds on $r$ and 
$|{\cal A}_{\rm CP}^{\rm dir}(B_d^0\to\pi^-K^+)|$ following from a 
measurement of the combined branching ratios (\ref{BR-char}) and 
(\ref{BR-neut}). In Section~\ref{discussion} we analyze the corresponding 
constraints arising from the present CLEO results and conclude our paper 
with a brief outlook in Section~\ref{conclusion}. 

\boldmath
\section{The description of $B^+ \to \pi^+K^0$ and
$B^0_d \to \pi^-K^+$ within the Standard Model}\label{formulae}
\unboldmath
Using a similar notation as in \cite{GHLR-ewp,ghlr}, the amplitudes for the 
decays under consideration can be written as \cite{PAPIII}
\begin{eqnarray}
A(B^+ \to \pi^+K^0) &=& {\cal P'} + c_d {\cal P}_{{\rm EW}}^{\prime{\rm C}} 
\label{ampl-char}  \\ 
A(B^0_d \to\pi^- K^+) &=& - \left[\left( P' + c_u P_{{\rm EW}}^{\prime{\rm 
C}}\right) + T'\right]\equiv - [ \tilde{P} + T']\,, \label{ampl-neut}
\end{eqnarray}
where ${\cal P'}$, $P'$ denote QCD penguin amplitudes, ${\cal P}_{{\rm 
EW}}^{\prime{\rm C}}$, $P_{{\rm EW}}^{\prime{\rm C}}$ correspond to
color-suppressed electroweak penguin contributions, and $T'$ is the 
color-allowed $\bar b\to\bar uu\bar s$ current-current amplitude\footnote{%
Let us note that we have neglected a highly suppressed annihilation 
contribution in (\ref{ampl-char}) which is expected to play an even less 
important role than the color-suppressed electroweak 
penguins \cite{GHLR-ewp}.}. The primes
remind us that we are dealing with $\bar b\to\bar s$ modes, the minus sign 
in (\ref{ampl-neut}) is due to our definition of meson states \cite{ghlr}, 
and $c_u=+2/3$ and $c_d=-1/3$ are the up- and down-type quark 
charges, respectively. The amplitude $\tilde{P}$ in (\ref{ampl-neut}) is a 
short-hand notation for the penguin contributions to $B^0_d \to\pi^- K^+$, 
i.e.\
\begin{equation}\label{Ptilde}
\tilde{P}\equiv P' + c_u P_{{\rm EW}}^{\prime{\rm C}}\,. 
\end{equation}
 
Whereas it is straightforward to show that the current-current amplitude $T'$
can be written in the Standard Model as 
\begin{equation}\label{tree}
T' = e^{i\gamma} e^{i\delta_{T'}} | T' |\,,
\end{equation}
where $\delta_{T'}$ is a CP-conserving strong phase and $\gamma$ the
usual angle of the unitarity triangle, the penguin amplitude 
$\tilde{P}$ is more involved. Here one has to deal with three different
contributions corresponding to penguins with internal up-, charm- and
top-quark exchanges. Taking into account all three of these contributions 
and {\it not} assuming dominance of internal top-quarks as is frequently 
done in the literature (we will comment on this point in 
Section~\ref{estimates}), it can be shown that the $\bar b\to\bar s$ 
penguin amplitude $\tilde{P}$ takes the following form~\cite{F97,fbf}:
\begin{equation}\label{pen}
\tilde{P} = e^{i\pi} e^{i\delta_{\tilde{P}}} | \tilde{P} |\,.
\end{equation}
Here $\delta_{\tilde{P}}$ is again a CP-conserving strong phase arising from
final state interactions, while only a trivial CP-violating weak phase 
appears in (\ref{pen}) as $e^{i\pi}=-1$. The amplitude structure of 
(\ref{ampl-char}) is analogous. 

Consequently the amplitude for the decay of the neutral $B^0_d$ meson is 
given by
\begin{equation}\label{ampl-n}
A(B^0_d \to \pi^- K^+) = e^{i\delta_{\tilde{P}}} | \tilde{P} | 
                       \left[1 - e^{i\gamma} e^{i\delta} r \right],
\end{equation}
where 
\begin{equation}\label{def-r}
r \equiv\frac{| T' |}{| \tilde{P} |}
\end{equation}
and $\delta$ is defined as the difference of the strong phases of $T'$ and 
$\tilde{P}$ through 
\begin{equation}\label{def-delta}
\delta \equiv \delta_{T'}- \delta_{\tilde{P}}\,. 
\end{equation}
Taking into account phase space, the branching ratio for $B^0_d \to \pi^- K^+$
is given by
\begin{equation}\label{br-neutral}
\mbox{BR}(B^0_d \to \pi^- K^+)=\frac{\tau_{B_d^0}}{16\pi M_{B_d^0}}
\Phi\left(M_{\pi^-}/M_{B_d^0},M_{K^+}/M_{B_d^0}\right)\,\left|A(B^0_d \to 
\pi^- K^+)\right|^2,
\end{equation}
where $\tau_{B_d^0}$ is the $B^0_d$ lifetime and 
\begin{equation}
\Phi(x,y)=\sqrt{\left[1-(x+y)^2\right]\left[1-(x-y)^2\right]}
\end{equation}
the usual two-body phase space function. Using (\ref{ampl-n}) gives
\begin{equation}\label{ampl-sq}
\left| A(B^0_d \to\pi^- K^+)\right|^2 = | \tilde{P} |^2 \left[1-2\,r 
\cos(\delta + \gamma) + r^2\right],
\end{equation}
while we have for the CP-conjugate process 
\begin{equation}\label{ampl-sqCP}
\left|A(\overline{B^0_d} \to \pi^+K^-)\right|^2 = | \tilde{P} |^2 \left[
1-2\,r \cos(\delta - \gamma) + r^2\right]
\end{equation}
corresponding to the replacement $\gamma\to-\gamma$.
The present data (\ref{BR-neutres}) reported recently by CLEO 
is an average over $B^0_d$ and $\overline{B^0_d}$ decays that is given by
\begin{equation}\label{untag-neutral}
\mbox{BR}(B_d \to \pi^\mp K^\pm)=\frac{\tau_{B_d}}{16\pi M_{B_d}}
\Phi\left(M_{\pi}/M_{B_d},M_{K_u}/M_{B_d}\right)\,\left\langle|A(B_d \to 
\pi^\mp K^\pm)|^2\right\rangle
\end{equation}
with
\begin{equation}
\left\langle|A(B_d\to\pi^\mp K^\pm)|^2\right\rangle\equiv\frac{1}{2}
\left(|A(B^0_d\to\pi^-K^+)|^2+|A(\overline{B^0_d}\to\pi^+K^-)|^2\right)\,.
\end{equation}
Combining (\ref{ampl-sq}) and (\ref{ampl-sqCP}) yields
\begin{equation}\label{ampl-av}
\left\langle|A(B_d\to\pi^\mp K^\pm)|^2\right\rangle=|\tilde{P}|^2\left[1-2\,r 
\cos\delta\,\cos\gamma + r^2\right],
\end{equation}
whereas the direct CP-violating asymmetry (\ref{def-CP}) can be expressed as
\begin{equation}\label{ACP}
{\cal A}_{\rm CP}^{\rm dir}(B_d^0\to\pi^-K^+)=2\,
\frac{|\tilde{P}|^2}{\left\langle|A(B_d
\to\pi^\mp K^\pm)|^2\right\rangle}\,r\,\sin\delta\,\sin\gamma\,.
\end{equation}
Taking into account that no non-trivial CP-violating weak phase is present 
in (\ref{ampl-char}) implies 
\begin{equation}\label{no-CP}
A(B^+\to\pi^+K^0)=A(B^-\to\pi^-\overline{K^0})\,,
\end{equation}
so that we get
\begin{equation}\label{av-br-char}
\mbox{BR}(B^\pm \to \pi^\pm K)=\frac{\tau_{B_u}}{16\pi M_{B_u}}
\Phi\left(M_{\pi}/M_{B_u},M_{K_d}/M_{B_u}\right)\,|A(B^+\to\pi^+ K^0)|^2\,.
\end{equation}

The color-suppressed electroweak penguin contributions ${\cal P}_{{\rm 
EW}}^{\prime{\rm C}}$ and $P_{{\rm EW}}^{\prime{\rm C}}$ in (\ref{ampl-char}) 
and (\ref{ampl-neut}) are expected to play a very minor role with respect 
to the QCD penguin amplitudes ${\cal P'}$ and $P'$ as we will see explicitly 
in Section~\ref{estimates} \cite{F97}. Neglecting these 
contributions and using the $SU(2)$ isospin symmetry of strong interactions 
allows us to relate the penguin amplitude $\tilde{P}$ relevant for 
$B_d\to\pi^\mp K^\pm$ to the $B^\pm\to\pi^\pm K$ decay amplitude through
\begin{equation}\label{Ptilde-det}
\tilde{P}=P'={\cal P'}=A(B^+\to\pi^+ K^0)\,.
\end{equation}
The magnitude of the right-hand side of this equation can be obtained 
from the measured $B^\pm\to\pi^\pm K$ branching ratio with the help of 
(\ref{av-br-char}). Consequently we get the following relation:
\begin{equation}\label{Simple-ACP}
{\cal A}_{\rm CP}^{\rm dir}(B_d^0\to\pi^-K^+)=2\,
\frac{\mbox{BR}(B^\pm\to\pi^\pm K)}{\mbox{BR}(B_d\to\pi^\mp K^\pm)}
\,r\,\sin\delta\,\sin\gamma\,,
\end{equation}
where the very small phase space difference between $B^\pm\to\pi^\pm K$
and $B_d\to\pi^\mp K^\pm$ has been neglected and the relevant $B$ lifetime
and mass ratios have been set to unity. 

\section{Semiquantitative estimates}\label{estimates}
Let us have a brief look at the theoretical framework to describe the 
$B\to\pi K$ decays relevant for our analysis. They are described by 
low energy effective Hamiltonians taking the following form:
\begin{eqnarray}
\lefteqn{\heff=}\label{ham}\\
&&\frac{G_{\rm F}}{\sqrt{2}}\left[V_{us}^\ast V_{ub}\sum\limits_{k=1}^2Q_k^u
C_k(\mu)+V_{cs}^\ast V_{cb}\sum\limits_{k=1}^2Q_k^c C_k(\mu)
-V_{ts}^\ast V_{tb}\sum\limits_{k=3}^{10}Q_k C_k(\mu)\right]\,,\nonumber
\end{eqnarray}
where $Q_k$ are local four-quark operators and $C_k(\mu)$ denote
Wilson coefficient functions calculated at a renormalization scale $\mu=
{\cal O}(m_b)$. The technical details of the evaluation of
such Hamiltonians beyond the leading logarithmic approximation has been
reviewed recently in \cite{bbl-rev}, where the exact definitions of the
current-current operators $Q_{1,2}^u$, $Q_{1,2}^c$, the QCD penguin
operators $Q_3,\ldots,Q_6$, the electroweak penguin operators
$Q_7,\ldots,Q_{10}$, and numerical values of their Wilson coefficients
can be found. Note that the $Q_{1,2}^c$ operators contribute to $B\to\pi K$
modes only through penguin-like matrix elements (see e.g.~\cite{rf1,rf2}) 
that are included by definition in the penguin amplitudes. A similar comment 
applies to effects of inelastic final state interactions that originate 
e.g.\ from the rescattering process $B^0_d\to\{D^+_sD^-\}\to\pi^-K^+$. In 
our notation these contributions are related to penguin-like matrix elements 
of the current-current operators and are also included in $P'$.

The color-allowed amplitude $\overline{T'}=e^{-2i\gamma}\,T'$
contributing to $\overline{B^0_d}
\to\pi^+K^-$ is related to hadronic matrix elements of 
the current-current operators $Q_1^u$ and $Q_2^u$ given by \cite{PAPIII}
\begin{eqnarray}
\langle K^-\pi^+|Q_1^u|\overline{B^0_d}\rangle&=&
\langle K^-\pi^+|(\bar s_{\alpha}
u_{\beta})_{\VmA}(\bar u_{\beta}b_{\alpha})_{\VmA}|\overline{B^0_d}
\rangle\label{e11a}\\
\langle K^-\pi^+|Q_2^u|\overline{B^0_d}\rangle&=&\langle K^-\pi^+|
(\bar su)_{\VmA}(\bar ub)_{\VmA}|\overline{B^0_d}\rangle\,,\label{e11b}
\eea
where $\alpha$ and $\beta$ denote $SU(3)_{\mbox{\scriptsize C}}$ color
indices and V$-$A refers to the Lorentz structure $\gamma_\mu(1-
\gamma_5)$. 
As in the case of $Q_{1,2}^c$, penguin-like matrix elements of the 
current-current operators $Q_{1,2}^u$ with up-quarks runing as a virtual 
particles in the loops \cite{rf1,rf2} contribute by definition to the penguin 
amplitudes $\overline{P'}=P'$, $\overline{P_{{\rm EW}}^{\prime{\rm C}}}=
P_{{\rm EW}}^{\prime{\rm C}}$ and not to $\overline{T'}$. Introducing 
non-perturbative $B$-parameters, (\ref{e11a}) and (\ref{e11b}) can be
written as
\begin{eqnarray}
\langle K^-\pi^+|Q_1^u(\mu)|\overline{B^0_d}\rangle&=&
\frac{1}{3}B_1(\mu){\cal F}\\
\langle K^-\pi^+|Q_2^u(\mu)|\overline{B^0_d}\rangle&=&B_2(\mu){\cal F}\,,
\end{eqnarray}
where ${\cal F}$ corresponds to the ``factorized'' matrix element 
$\langle K^-|(\bar su)_{\VmA}|0\rangle\langle\pi^+|
(\bar ub)_{\VmA}|\overline{B^0_d}\rangle$ and $B_k (\mu) \neq 1$ parametrizes
deviations from factorization.
Consequently we get
\begin{equation}\label{CC-ampl}
\overline{T'}=-\frac{G_{\rm F}}{\sqrt{2}}V_{us}^\ast V_{ub}\left[
\frac{1}{3}\frac{B_1(\mu)}{B_2(\mu)}C_1(\mu)+C_2(\mu)\right]\,B_2(\mu)\,
{\cal F}\,,
\end{equation}
which can be written by introducing the phenomenological color-factor
$a_1$ \cite{BSW,nrsx} as
\begin{equation}\label{fact1}
\overline{T'}=-\frac{G_{\rm F}}{\sqrt{2}}V_{us}^\ast V_{ub}\,a_1\,{\cal F}\,.
\end{equation}
The minus sign is due to our definition of meson states (see also the 
remark after (\ref{ampl-neut})). 

For the following discussion $|T'|$ plays an important role and can be
written with the help of the ``factorization'' assumption as
\begin{equation}\label{fact}
\left.|T'|\right|_{{\rm fact}}=
\frac{G_{\rm F}}{\sqrt{2}}\,\lambda\,|V_{ub}|\,a_1\,\left(M_{B_d}^2-
M_\pi^2\right)\,f_{K}\,F_{B\pi}(M_K^2;0^+)\,,
\end{equation}
where $\lambda=0.22$ is the Wolfenstein parameter \cite{wolf} and ${\cal F}$ 
has been expressed in terms of quark-current form factors \cite{BSW}. The
presently allowed range for $|V_{ub}|$ is given by $(3.2\pm0.8)\cdot10^{-3}$
\cite{bf-rev}. Data from $\overline{B^0_d}\to D^{(\ast)+}\pi^-$ and 
$\overline{B^0_d}\to D^{(\ast)+}\rho^-$ decays imply $a_1=1.06\pm0.03\pm0.06$  
\cite{a1a2-exp}. From a theoretical point of view, $a_1$ is very stable for
$B$ decays and lies within the range $a_1=1.01\pm0.02$ \cite{buras}.
Although the ``factorization'' hypothesis \cite{facto} is in general
questionable, it may work with reasonable accuracy for the color-allowed 
current-current amplitude $T'$ \cite{bjorken}. Using the formfactor
$F_{B\pi}(M_K^2;0^+)=0.3$ as obtained in the BSW model \cite{BSW} yields
\begin{equation} \label{Tprime}
\left.|T'|\right|_{{\rm fact}}=a_1\cdot\left[\frac{|V_{ub}|}{3.2\cdot10^{-3}}
\right]\cdot7.8\cdot10^{-9}\,\mbox{GeV}\,.
\end{equation}

\begin{figure}   
\epsfysize=9.8cm
\centerline{\epsffile{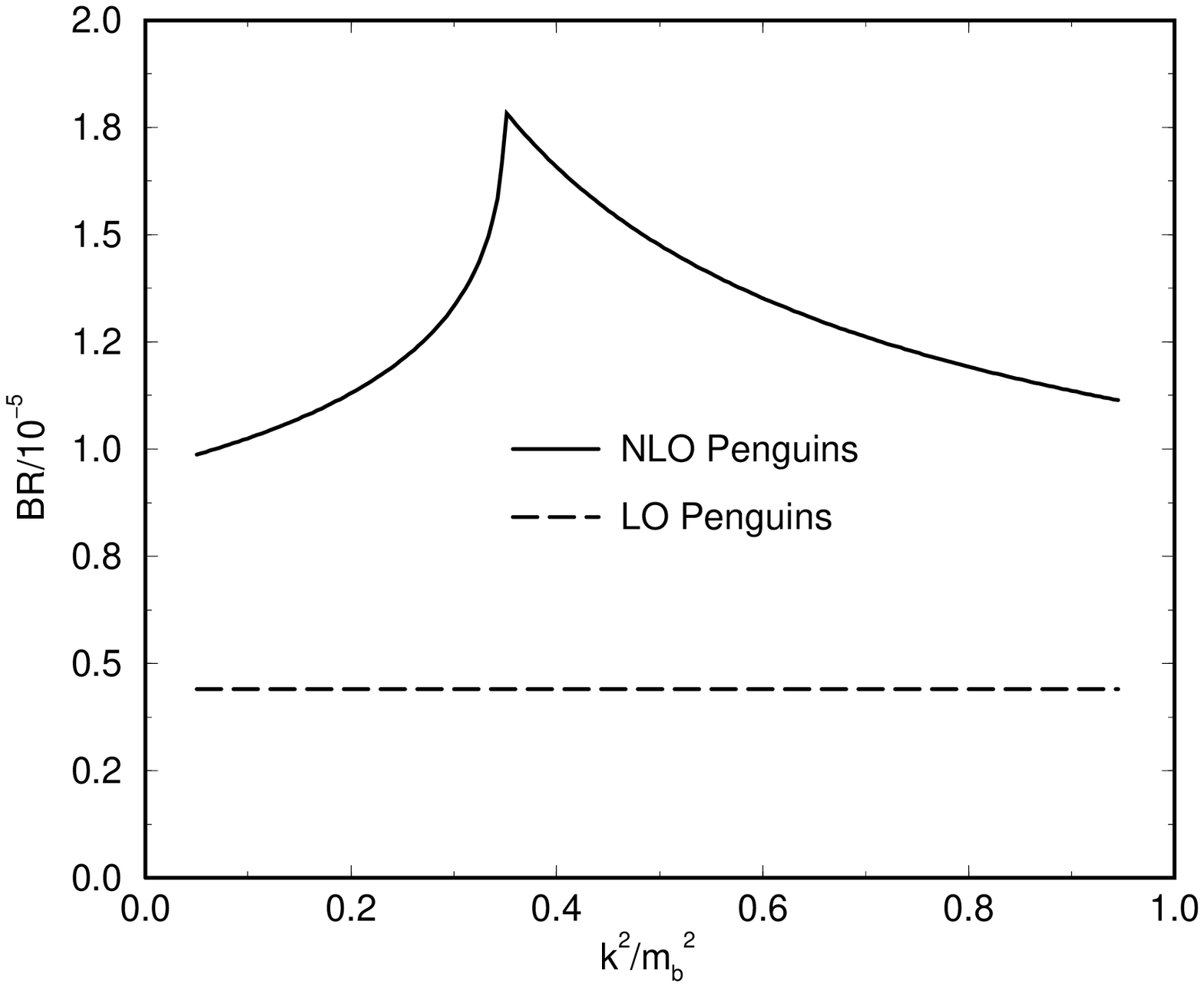}}
\caption[]{The dependence of BR$\left.(B^\pm\to\pi^\pm K)\right|_{\rm fact}$ 
on $k^2$ for $\Lambda_{\overline{{\rm MS}}}^{(4)}=0.3\,\mbox{GeV}$. The
difference between the NLO and LO curves is explained in the text.}
\label{fig:BR-est}
\end{figure}

In contrast to the case of $T'$, the use of the factorization assumption 
is questionable for the penguin amplitude $\tilde P$. Let us nevertheless
use that approach to get some feeling for the expected orders of magnitudes. 
Following the formalism developed in \cite{rf1,rf2,kps} and using the 
Wolfenstein expansion \cite{wolf} with $A=0.810\pm0.058$ and 
$R_b\equiv|V_{ub}|/(\lambda|V_{cb}|)=0.363\pm0.073$ \cite{al}, 
the $\bar b\to\bar s$ penguin amplitude (\ref{Ptilde}) can be expressed as
\begin{eqnarray}
\lefteqn{\left.\tilde P\right|_{\rm fact}=
-\frac{G_{\rm F}}{\sqrt{2}}A\lambda^2
\left[\frac{1}{3}\overline{C}_3+\overline{C}_4+\frac{1}{3}\overline{C}_9+
\overline{C}_{10}+\frac{2M_{K}^2}{m_sm_b}\left(\frac{1}{3}\overline{C}_5+
\overline{C}_6+\frac{1}{3}\overline{C}_7+\overline{C}_8\right)
\right.}\nonumber\\
&&\left.+\frac{\alpha_s(m_b)}{9\pi}\left\{\frac{10}{9}-G(m_c,k,m_b)\right\}
\left(1+\frac{2M_K^2}{m_sm_b}\right)\left\{\overline{C}_2+
\frac{\alpha_{\rm QED}}{\alpha_s(m_b)}\left(\overline{C}_1+\frac{1}{3}
\overline{C}_2\right)\right\}\right]\,{\cal F}\,,\label{pen-fact}
\end{eqnarray}
where we have used in addition to the factorization approximation 
the equations of motion for the quark fields leading to the terms 
proportional to $M_K^2/(m_sm_b)$.  The $\overline{C}_k$'s refer to $\mu=m_b$
and denote the next-to-leading order scheme-independent Wilson coefficient 
functions introduced by Buras et al.\ in \cite{Buras-NLO}. The function 
$G(m_c,k,m_b)$ is related to one-loop penguin matrix elements of the 
current-current operators $Q_{1,2}^c$ with internal charm-quarks and is 
given by 
\begin{equation}
G(m_c,k,m_b)=-\,4\int\limits_{0}^{1}\mbox{d}x\,x\,(1-x)\ln\left[\frac{m_c^{2}-
k^{2}\,x\,(1-x)}
{m_b^{2}}\right],
\end{equation}
where $m_c$ is the charm-quark mass and $k$ denotes some average 
four-momentum of the virtual gluons and photons appearing in corresponding 
penguin diagrams \cite{rf1,rf2}. Simple kinematical considerations at
the quark-level imply the following ``physical'' range for this 
parameter \cite{dt,gh}:
\begin{equation}
\frac{1}{4}\leqsim \frac{k^2}{m_b^2}\leqsim\frac{1}{2}\,.
\end{equation}
In the case of the $b\to s$ penguin processes considered here, 
the penguin-like matrix elements of $Q_{1,2}^u$ are highly suppressed
with respect to those of $Q_{1,2}^c$ by the CKM factor $|V_{us}^\ast V_{ub}|/
|V_{cs}^\ast V_{cb}|=\lambda^2R_b={\cal O}(0.02)$ and have been neglected in
(\ref{pen-fact}). These terms may lead to CP asymmetries in $B^\pm\to
\pi^\pm K$ that are at most of ${\cal O}(1\%)$ and consequently affect the 
relation (\ref{no-CP}) to a very small extent \cite{rf1,kps}.

\begin{figure}   
\epsfysize=9.8cm
\centerline{\epsffile{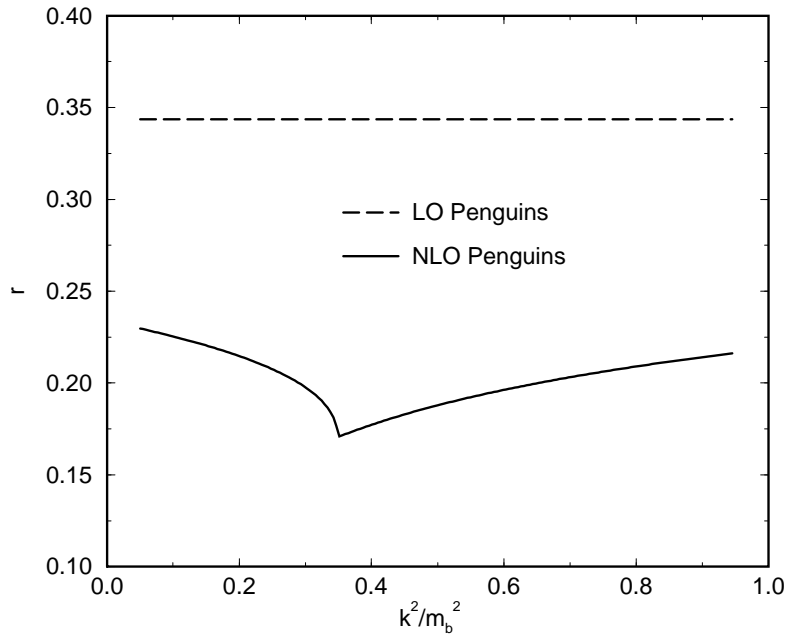}}
\caption[]{The dependence of $\left.r\right|_{\rm fact}$ on $k^2$ for 
$\Lambda_{\overline{{\rm MS}}}^{(4)}=0.3\,\mbox{GeV}$. The
difference between the NLO and LO curves is explained in the text.}
\label{fig:r-est}
\end{figure}

As was stressed in \cite{rf1}, a consistent calculation using next-to-leading 
order Wilson coefficients requires the inclusion of the penguin-like matrix 
elements of the current-current operators discussed above. The point is that 
the renormalization scheme dependences of these matrix elements cancel those 
of the $C_k$'s leading to the scheme independent Wilson coefficients 
$\overline{C}_k$. On the other hand, using leading order Wilson coefficients 
$C_k^{\rm LO}$, these matrix elements have to be dropped so that we have in 
this case
\begin{eqnarray}
\lefteqn{\left.\tilde P\right|_{\rm fact}^{\rm LO}=
-\frac{G_{\rm F}}{\sqrt{2}}A\lambda^2}\label{pen-simple}\\
&&\times\left[\frac{1}{3}C^{\rm LO}_3+C^{\rm LO}_4+\frac{1}{3}C^{\rm LO}_9+
C^{\rm LO}_{10}+\frac{2M_{K}^2}{m_sm_b}\left(\frac{1}{3}C^{\rm LO}_5+
C^{\rm LO}_6+\frac{1}{3}C^{\rm LO}_7+C^{\rm LO}_8\right)
\right]\,{\cal F}\,.\nonumber
\end{eqnarray}

Using numerical values for the Wilson coefficients, we find that the 
contribution of the electroweak penguin operators to $\tilde P$ is below 
the ${\cal O}(1\%)$ level so that the approximation of neglecting the
$P_{{\rm EW}}^{\prime{\rm C}}$ contributions (see the comment before 
(\ref{Ptilde-det})) seems to be on solid ground. Evaluating the branching 
ratios for the penguin
mode $B^\pm\to\pi^\pm K$ corresponding to (\ref{pen-fact}) and 
(\ref{pen-simple}), we find (as can be seen already in the tables given 
in \cite{rf1}; see also \cite{cfms}) that the penguins with internal 
charm-quarks lead to a dramatic enhancement. This feature can be seen 
in Fig.~\ref{fig:BR-est}, where we show the dependence of 
BR$\left.(B^\pm\to\pi^\pm K)\right|_{\rm fact}$ on $k^2$ for $A=0.81$,
$F_{B\pi}(M_K^2;0^+)=0.3$ and $\tau_{B_u}=1.6\,\mbox{ps}$. Using 
(\ref{fact1}) with $a_1=1$, these branching ratios correspond to the 
amplitude ratios $r$ shown in Fig.~\ref{fig:r-est}, where we have chosen
$R_b=0.36$ to evaluate these plots. Consequently in this rather simple
model calculation the penguin matrix elements with internal charm-quarks 
lead to an enhancement of BR$(B^\pm\to\pi^\pm K)$ by a factor of ${\cal O}(3)$
and to a reduction of $r$ by ${\cal O}(2)$. Interestingly the CLEO result
(\ref{BR-charres}) does already rule out the LO curve in Fig.~\ref{fig:BR-est}.
The NLO result, however, still has some dependence on $k^2$ which will
disappear once a nonperturbative calculation of the matrix elements becomes 
available. Still the agreement with the present CLEO data is remarkable
although it is a bit on the lower side. 

In a similar spirit we can arrive at some estimate of the 
CP-conserving strong phase defined in (\ref{def-delta}). 
We obtain $\delta=0^\circ$ if we use (\ref{pen-simple}) for $\tilde P$.
Including the important penguin matrix elements with internal charm-quarks
through (\ref{pen-fact}) gives values of $\delta$ within the range 
$-30^\circ\leqsim\delta\leqsim0^\circ$. In spite of all the caveats connected 
with factorization we still consider it safe to extract the sign of 
$\cos \delta$ from this discussion. Hence we have very probably 
$\cos\delta>0$.

Finally we want to stress that none of the crude estimates discussed 
above are needed for the analysis presented in section \ref{discussion}. 
The only purpose to include these results in our paper is to update 
previous theoretical work given the new input from CLEO.  

\boldmath
\section{Constraints on $\gamma$, $r$ and $|{\cal A}_{\rm CP}^{\rm dir}
(B_d^0\to\pi^-K^+)|$}
\unboldmath
\label{bounds}
In this section we will derive some simple relations allowing to 
constrain the CKM angle $\gamma$ by measuring the combined branching
ratios BR$(B^\pm\to\pi^\pm K)$ and BR$(B_d\to\pi^\mp K^\pm)$ specified in
(\ref{BR-char}) and (\ref{BR-neut}), respectively. Such measurements 
allow us moreover to restrict the range of $r$ and to give upper bounds for 
the direct CP asymmetry $|{\cal A}_{\rm CP}^{\rm dir}(B_d^0\to\pi^-K^+)|$. 
To this end the quantity
\begin{equation}\label{def-R}
R\equiv\frac{\left\langle|A(B_d\to\pi^\mp K^\pm)|^2\right\rangle}{|\tilde 
P|^2}
\end{equation}
turns out to be very useful. Neglecting the small phase space difference
between $B_d\to\pi^\mp K^\pm$ and $B^\pm\to\pi^\pm K$ and using 
(\ref{untag-neutral}), (\ref{av-br-char}) and (\ref{Ptilde-det}) yields
\begin{equation}\label{R-det}
R=\frac{\mbox{BR}(B_d\to\pi^\mp K^\pm)}{\mbox{BR}(B^\pm\to\pi^\pm K)}\,,
\end{equation}
where the ratio of the relevant $B$-meson lifetimes and masses has been 
set to unity as in (\ref{Simple-ACP}). 
Consequently $R$ can be fixed through the measured branching ratios 
(\ref{BR-charres}) and (\ref{BR-neutres}). Using (\ref{ampl-av}) gives
\begin{equation}\label{A-def}
C\equiv\cos\delta\,\cos\gamma=\frac{1-R}{2r}+\frac{1}{2}r\,.
\end{equation}
In the following considerations we will keep $\delta$ as a free parameter 
leading to the relation $|\cos\gamma|\geq|C|$ which implies
\begin{equation}\label{GAMMA0}
\gamma_0=\arccos(|C|)
\end{equation}
for the range (\ref{gamma-bound1}). Since $C$ is given by the product 
of two cosines, it has to lie within the range $-1\leq C\leq+1$. 
As $R$ is fixed through (\ref{R-det}), this range has the following 
implication for $r$:
\begin{equation}\label{range-r}
\left|1-\sqrt{R}\right|\leq r\leq1+\sqrt{R}\,.
\end{equation}
The magnitude of the direct CP-violating asymmetry (\ref{ACP}) in 
$B_d^0\to\pi^-K^+$ can be expressed with the help of $R$ and $C$ as
\begin{equation}
\left|{\cal A}_{\rm CP}^{\rm dir}(B_d^0\to\pi^-K^+)\right|=2\,
\frac{r}{R}\,\sqrt{\sin^2\gamma-C^2\tan^2\gamma}\,.
\end{equation}
Keeping $r$ and $R$ fixed, this CP asymmetry takes its maximal value
\begin{equation}\label{ACPmax}
\left|{\cal A}_{\rm CP}^{\rm dir}(B_d^0\to\pi^-K^+)\right|_{\rm 
max}=2\,\frac{r}{R}\,\left(1-|C|\right)
\end{equation}
for 
\begin{equation}
\gamma_{\rm max}=\arccos\left(\sqrt{|C|}\right)\quad
\lor\quad\gamma_{\rm max}=180^\circ - \arccos\left(\sqrt{|C|}\right)
\, ,
\end{equation}
where $C$ is expressed in terms of $r$ and $R$ in (\ref{A-def}).

\section{Implications of the CLEO measurements}\label{discussion}
In this section we shall discuss the implications of the recent CLEO 
measurements using the relations derived in the last section. In particular 
it will become clear that an experimental improvement will make these 
constraints much more stringent as they appear using present data. 

The recent CLEO measurements given in (\ref{BR-charres}) and (\ref{BR-neutres})
allow to determine the value of $R$. Putting the numbers and adding the 
errors in quadrature gives
\begin{equation}\label{R-exp}
R = 0.65 \pm 0.40\,.
\end{equation}
In Fig.~\ref{fig:A} we show the dependence of the quantity $C$ definded
by (\ref{A-def}) on the amplitude ratio $r=|T'|/|\tilde P|$ for various 
values of $R$ within that experimentally fixed range. This figure illustrates 
nicely the constraints on $r$ given in (\ref{range-r}) arising from the fact 
that $C$ has to lie within the range $-1\leq C\leq+1$.

\begin{figure}   
\epsfysize=9.8cm
\centerline{\epsffile{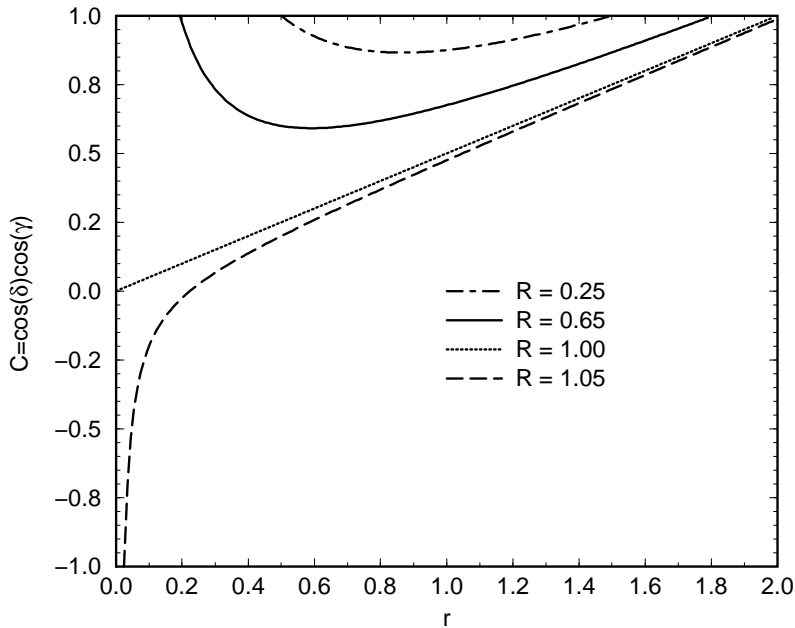}}
\caption[]{The dependence of $C$ on the amplitude
ratio $r$ for various values of $R$.}
\label{fig:A}
\end{figure}

As is obvious from the discussions in Sections~\ref{formulae} and 
\ref{estimates}, $r$ can in 
principle be determined in a direct way. Using (\ref{Ptilde-det}), the 
denominator is 
fixed by the penguin decay $B^\pm \to \pi^\pm K$. The numerator is more 
difficult to obtain. One possibility is to use (\ref{Tprime}) leading to
\begin{equation}
r=0.16\cdot a_1\cdot\left[\frac{|V_{ub}|}{3.2\cdot10^{-3}}\right]
\sqrt{\left[\frac{2.3\cdot10^{-5}}{\mbox{BR}(B^\pm \to \pi^\pm K)}\right]
\cdot\left[\frac{\tau_{B_u}}{1.6\,\mbox{ps}}\right]}\,.
\end{equation}
However, this expression relies on factorization and uncertainties become hard 
to estimate. Another 
way would be to relate $|T'|$ to some current-current dominated process. 
Assuming flavor $SU(3)$, a possible mode is $B^\pm \to \pi^\pm \pi^0$ which 
receives only color-allowed and color-suppressed current-current and 
negligibly small electroweak penguin contributions. Including factorizable 
$SU(3)$-breaking we obtain \cite{PAPIII}
\begin{equation}\label{Tprime-det}
|T'| = \lambda\,\frac{f_K}{f_\pi}\,\sqrt{2}\,|A (B^\pm \to \pi^\pm \pi^0)|\,,
\end{equation}
where we have neglected the color-suppressed current-current contributions.
An interesting experimental consistency check would be the comparison of 
(\ref{Tprime}) with (\ref{Tprime-det}). Unfortunately the $B^\pm\to\pi^\pm 
\pi^0$ mode has not yet been measured. However, recently the CLEO 
collaboration has reported the following upper limit for the corresponding
branching ratio~\cite{CLEO}:
\begin{equation}\label{Bpipi-limit}
BR(B^\pm\to\pi^\pm\pi^0)<2.0\cdot10^{-5}\,,
\end{equation}
which can be translated easily into an upper bound on $r$. Using the minimal 
and central values $1.1\cdot10^{-5}$ and $2.3\cdot10^{-5}$ of the branching 
ratio in (\ref{BR-charres}), (\ref{Tprime-det}) leads to 
\begin{equation}\label{r-limits}
r\,\leqsim\,0.51\quad\mbox{and}\quad r\,\leqsim\,0.35\,,
\end{equation}
respectively. However, given all the assumptions leading to these bounds, 
we shall not exclude the possibility of having a larger value of $r$ 
within the limits (\ref{range-r})
in what follows and consider all quantities as functions of $r$. 

\begin{figure}   
\epsfysize=9.8cm
\centerline{\epsffile{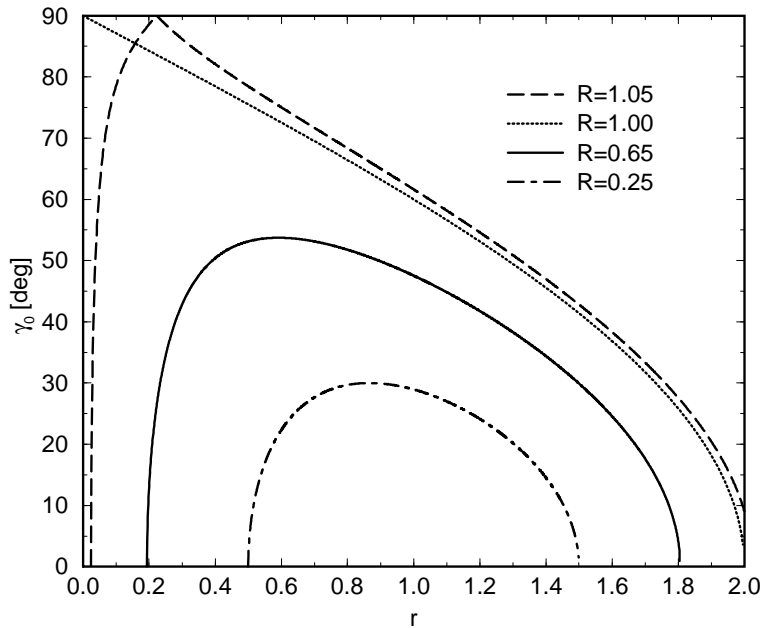}}
\caption[]{The dependence of $\gamma_0$ constraining the CKM angle $\gamma$
through (\ref{gamma-bound1}) on the amplitude ratio $r$ for various values 
of $R$.}
\label{fig:gamma}
\end{figure}

In Fig.~\ref{fig:gamma} we plot the constraint on $\gamma$ as a function 
of $r$. As usual we shall assume that $\gamma$ ranges between $0^\circ$ 
and $180^\circ$ which is determined from CP violation in the Kaon 
system~\cite{bf-rev}. Due to the two possibilities for the sign of 
$\cos \delta$ we have to discuss two cases. For positive $\cos \delta$ 
the sign of $\cos \gamma$ is the same as the one of $C$. Thus for $R<1$ 
we can constrain the angle $\gamma$ between
$0^\circ$ and $\gamma_0 < 90^\circ$. For the small window $R>1$, which 
is still allowed due to the large experimental uncertainty, $C$ becomes 
negative for $r<\sqrt{R-1}$ implying that $\gamma$ lies within the range 
$90^\circ < 180^\circ - \gamma_0 \le\gamma\le 180^\circ$ in that case. 
If $\cos \delta$ is negative, the situation reverses. 

It is obvious that the constraint on $\gamma$ becomes more restrictive the 
smaller the value of $R$ is. As can be seen from Fig.~\ref{fig:gamma}, 
$R = 1$ is an important special case. The point is that for $R < 1$ one can 
always constrain $\gamma$ independent of $r$, while $R>1$ requires some 
knowledge about $r$. For $R<1$ the maximal value of $\gamma_0$ is given by
\begin{equation}
\gamma_0^{\rm max} = \arccos \left( \sqrt{1-R} \right) .
\end{equation} 
In particular, if $R$ is significantly smaller than one, we may place 
stringent restrictions on $\gamma$. For instance, taking the central value 
of the CLEO measurement we have $\gamma_0^{\rm max} = 54^\circ$; for a 
value at the lower end of (\ref{R-exp}) we have even $\gamma_0^{\rm max} = 
30^\circ$. If one is to take the lower limit in (\ref{r-limits}) corresponding
to $R=0.65$ serious, one finds $\gamma_0 < 48^\circ$.

In Fig.~\ref{fig:ACP} we show the dependence of
$\left|{\cal A}_{\rm CP}^{\rm dir}(B_d^0\to\pi^-K^+)\right|$ 
on $r$ for various values of $R$ within the 
experimental range (\ref{R-exp}). In contrast to the case of $\gamma$ it is 
impossible to constrain that CP asymmetry without any knowledge of $r$. 
Coming back to the previous example, we have  
$\left|{\cal A}_{\rm CP}^{\rm dir}(B_d^0\to\pi^-K^+)\right| \le 0.35$
for $R = 0.65$ and $r$ bounded by the lower value of~(\ref{r-limits}). 

\begin{figure}   
\epsfysize=9.8cm
\centerline{\epsffile{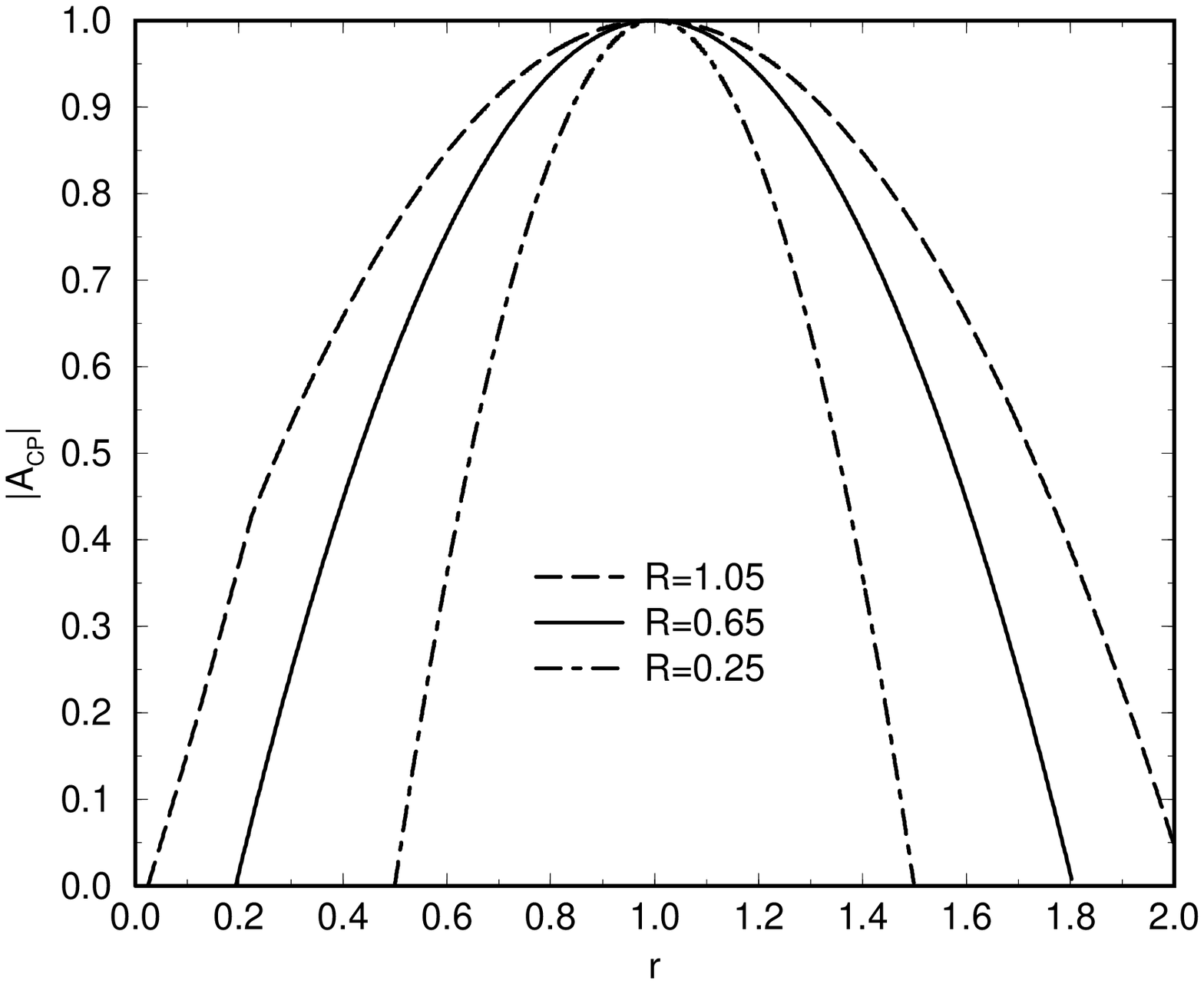}}
\caption[]{The dependence of the maximal value (\ref{ACPmax}) of 
$\left|{\cal A}_{\rm CP}^{\rm dir}(B_d^0\to\pi^-K^+)\right|$ on the amplitude
ratio $r$ for various values of $R$.}
\label{fig:ACP}
\end{figure}

\section{Conclusions and outlook}\label{conclusion}
In the present paper we have shown that a measurement of the combined 
branching ratios (\ref{BR-char}) and (\ref{BR-neut}) allows to obtain useful 
constraints on $\gamma$ and direct CP violation in $B_d \to \pi^\mp K^\pm$
even with rather large experimental uncertainties. Needless to say, an 
improvement on the experimental side will sharpen the bounds on $\gamma$ 
substantially, in particular it would be useful to further constrain $R$ 
to the region $R<1$. Obviously another important step would be a separate
measurement of the $B^+\to\pi^+K^0$, $B^-\to\pi^-\overline{K^0}$ and
$B^0_d\to\pi^-K^+$, $\overline{B^0_d}\to\pi^+K^-$ branching ratios which
may lead to a determination of $\gamma$ as proposed in \cite{PAPIII}.

Looking at the bounds on $\gamma$ we have derived in the present paper, they 
are complementary to what is obtained from a global fit of the 
unitarity triangle using experimental data on $|V_{cb}|$, 
$|V_{ub}|/|V_{cb}|$, $B^0_d$--$\overline{B^0_d}$ mixing and CP violation in 
the neutral $K$-meson system~\cite{bf-rev,al}. 
Typically that  range for $\gamma$ using present data is
\begin{equation}\label{gamma-normal}
40^\circ \leqsim \gamma \leqsim 140^\circ\,.
\end{equation}
Note that the allowed range is symmetric around $\gamma= 90^\circ$, while in 
our approach we {\it exclude} a range symmetric with respect to $90^\circ$
for $R<1$. For instance, taking the central value $R = 0.65$, we have 
$0 \le \gamma \le \gamma_0^{\rm max} = 54^\circ$ or $126^\circ \le \gamma 
\le 180^\circ$ depending on the sign of $\cos \delta$. In order to be 
compatible, this means that $\gamma$ has to be either between $40^\circ$ 
and $54^\circ$ or between $126^\circ$ and $140^\circ$. Based on the 
discussion of Section~\ref{estimates} we conclude that  the former range 
is the preferred one since most probably $\cos\delta>0$. In the case of the
central value of (\ref{gamma-normal}), i.e.\ $\gamma=90^\circ$, we have
$C=0$ and get therefore the relation $r=\sqrt{R-1}$ between $r$ and $R$
independent of the value of $\delta$. Note that in this case necessarily 
$R>1$. Using our bound (\ref{r-limits}) on
$r$ implies thus $1<R\,\leqsim\,1.25$ for $\gamma=90^\circ$. If $r$ should 
be of ${\cal O}(0.2)$ as expected, we would practically fix $R$ to be 
$1<R\,\leqsim\,1.04$. However, this corresponds to the upper end of
the present CLEO range. If $\gamma$ is close to $90^\circ$, 
future measurements either have a value of $R$ close to unity or 
it will become increasingly difficult to accommodate the situation 
within the Standard Model.

Although some of our bounds are independent of the ratio $r$, this quantity
is still one of the main ingredients of the presented approach. The range 
implied by 
pure consistency given in (\ref{range-r}) is quite generous. Using other 
input to access $|T'|$ and $|\tilde{P}|$, such as factorization for the 
color-allowed current-current amplitude or data on $B^\pm \to \pi^\pm \pi^0$, 
consistently indicates small values of $r$. It is interesting to note that 
these smallish values are already at the edge of compatibility with the CLEO 
measurements. 

Another important experimental task is to search for direct CP violation
in $B_d\to\pi^\mp K^\pm$ which would immediately rule out ``superweak''
models of CP violation \cite{s-weak}. Ruling out these scenarios with the 
help of that CP asymmetry is, however, not the only possibility; if one should 
measure CP-violating effects in $B_d\to\pi^\mp K^\pm$ that are inconsistent
with the upper limits on $\left|{\cal A}_{\rm CP}^{\rm dir}
(B_d^0\to\pi^-K^+)\right|$ obtained along the lines proposed in our paper
one would also have indications for physics beyond the Standard Model. 

In conclusion, we have demonstrated that the combined $B\to\pi K$ branching
ratios reported recently by the CLEO collaboration may lead to stringent
constraints on $\gamma$ that are complementary to the presently allowed
region of that angle obtained with the help of the usual indirect methods
to determine the unitarity triangle. These measurements provide in
addition a powerful tool to check the consistency of the Standard Model
description of these decays and to search for ``New Physics''. In this
respect direct CP violation in $B_d\to\pi^\mp K^\pm$ is also expected to
play an important role. Once more data come in confirming values of $R<1$,
the $B\to\pi K$ modes discussed in our paper may put 
the Standard Model to a decisive test and could open a window to 
``New Physics''.

\section*{Acknowledgments}

This work was supported by DFG under contract Ma 1187/7-1,2
and by the Graduierten\-kolleg ``Elementarteilchenphysik and Beschleunigern''. 
R.F. would like to thank James Alexander and Frank W\"urthwein for 
conversations about the $B\to\pi K$ CLEO results.

\end{document}